\documentstyle[aps,prb,multicol]{revtex}

\begin{document}

\title{Strong correlation effects of the Re 5$d$ electrons on the metal-insulator transition in Ca$_2$FeReO$_6$}

\author{H. Iwasawa, T. Saitoh,\cite{adrT} Y. Yamashita, and D. Ishii}
\address{Department of Applied Physics, Tokyo University of Science, Shinjuku-ku, Tokyo 162-8601, Japan}

\author{H. Kato\cite{adrKato}}
\address{Correlated Electron Research Center (CERC), National Institute of Advanced Industrial Science and Technology (AIST), Tsukuba 305-8562, Japan}

\author{N. Hamada}
\address{Department of Physics, Tokyo University of Science, Chiba 278-8510, Japan}

\author{Y. Tokura}
\address{Correlated Electron Research Center (CERC), National Institute of Advanced Industrial Science and Technology (AIST), Tsukuba 305-8562, Japan; \\Spin Superstructure Project, ERATO, Japan Science and Technology Corporation (JST), Tsukuba 305-8562, Japan; \\and Department of Applied Physics, University of Tokyo, Tokyo 113-8656, Japan}

\author{D. D. Sarma\cite{adrDD}}
\address{Solid State and Structural Chemistry Unit, Indian Institute of Science, Bangalore 560 012, India}

\date{Received \today}

\maketitle

\begin{abstract}

	We have investigated the electronic structure of polycrystalline Ca$_2$FeReO$_6$ using photoemission spectroscopy and band-structure calculations within the local-density approximation+$U$ (LDA+$U$) scheme. In valence-band photoemission spectra, a double-peak structure which is characteristic of the metallic double perovskite series has been observed near the Fermi level ($E_{\rm F}$), although it is less distinct compared to the Sr$_2$FeMoO$_6$ case. The leading near-$E_{\rm F}$ structure has a very weak spectral weight at $E_{\rm F}$ above the metal-insulator transition (MIT) temperature $T_{\rm MI}$ of $\sim$140 K, and it loses the $E_{\rm F}$ weight below $T_{\rm MI}$, forming a small energy gap.
	To reproduce this small energy gap in the calculation, we require a very large effective $U$ ($U_{\rm eff}$) for Re (4 eV) in addition to a relatively large $U_{\rm eff}$ for Fe (4 eV). Although the most of the experimental features can be interpreted with the help of the band theory, the overall agreement between the theory and the experiment was not satisfactory. We demonstrate that the effective transfer integral between Fe and Re is actually smaller than that between Fe and Mo in Ca$_2$FeMoO$_6$, which can explain both MIT and very high ferrimagnetic transition temperature.

\end{abstract}


 
\vspace*{-0.3 in}

\begin{multicols}{2}

\section{Introduction}\label{secIntr}

	Ordered double perovskite-type oxides $A_2$$B$$B^{\prime}$O$_6$ have a natural super-lattice structure in which the transition metal sites $B$ and $B^{\prime}$ are alternatively located at the $B$ site of the simple perovskite structure $AB$O$_3$. Pioneering studies of ordered double perovskites have been already done in the 1960s.\cite{Longo,Sleight,Patterson,Galasso,Sleight2}
	Several decades after, the discovery of a large room temperature (RT) tunneling magnetoresistance (TMR) in a low magnetic field in Sr$_2$FeMoO$_6$ and Sr$_2$FeReO$_6$ triggered revived interest in such ordered double perovskites in view of the possibility of industrial applications for spin-electronic devices.\cite{KobayashiNature,MoritomoPRBR,KobayashiRe,Wolf} These materials were predicted to have half-metallic density of states (DOS) like the manganites.\cite{KobayashiNature,KobayashiRe} Several band-structure calculations have shown that metallic ordered double perovskites are generally half-metallic,\cite{Moritomo2,Sarma,Fang,Wu,Kang,Tom,Szotek} and the half metallic DOS has been confirmed by optical and valence-band photoemission studies.\cite{Kang,Tom,Tomioka} 

	Among the broad class of the ordered double perovskites, (Sr$_{\rm 1-y}$Ca$_{\rm y}$)$_2$FeReO$_6$ is known as an exceptional system.\cite{Kato,Philipp} In particular, Ca$_2$FeReO$_6$ shows metal-insulator transition (MIT) as well as has a very high ferrimagnetic transition temperature ($T_{\rm c}$) of 540K,\cite{Kato,Westburg} which is the second highest in the known ordered double perovskites.\cite{Kato2}
	Kato {\it et al.}\cite{Kato} have studied the temperature and composition ($y$) dependent MIT in (Sr$_{\rm 1-y}$Ca$_{\rm y}$)$_2$FeReO$_6$; the ground state changes from metallic to insulating at $y$$\sim$0.4 with increasing $y$, and the transition temperature ($T_{\rm MI}$) rises up to $\sim$150 K at the Ca end. They have also reported that this MIT accompanies a structural phase transition.
	Since Ca$_2$FeReO$_6$ and Sr$_2$FeReO$_6$ (or Ba$_2$FeReO$_6$) have the same electronic configuration Fe$^{3+}$(3$d$$^5$; $t^{3}_{2g\uparrow}$$e^{2}_{g\uparrow}$)$-$Re$^{5+}$(5$d$$^2$; $t^{2}_{2g\downarrow}$) within the simple ionic picture, the variation of the bond-angle distortion from $\sim$$180^\circ$ to $\sim$$152^\circ$ due to Ca substitution for Sr (or Ba) alone seems to be able to account for the insulating ground state of Ca$_2$FeReO$_6$ at a first glance.\cite{Gopala}
	It is of course true that the effective transfer integral between Fe and Re via oxygen and hence the one-electron bandwidth are expected to decrease with increasing bond-angle distortion. However, Ca$_2$FeMoO$_6$ and Sr$_2$FeMoO$_6$, which have the same Fe$^{3+}$(3$d$$^5$; $t^{3}_{2g\uparrow}$$e^{2}_{g\uparrow}$)$-$Mo$^{5+}$(4$d$$^1$; $t^{1}_{2g\downarrow}$) configuration, are both metallic although the Fe-O-Mo angles are $\sim$$153^\circ$ and $\sim$$180^\circ$, respectively.\cite{Alonso,Ritter} It is, therefore, difficult to attribute the insulating nature of Ca$_2$FeReO$_6$ solely to the bond-angle distortion.

	Recently, Oikawa {\it et al.} have studied the crystal structure and magnetic properties of Ca$_2$FeReO$_6$ using neutron powder diffraction (NPD) at RT and 7 K.\cite{Oikawa} They have observed a distinct change of the crystal structure from a monoclinic structure in the ferrimagnetic metallic (FM) phase above 140 K, namely $T_{\rm MI}$, to another monoclinic structure with different monoclinic angles in the ferrimagnetic insulating (FI) phase below $T_{\rm MI}$. Crossing $T_{\rm MI}$, FeO$_6$ and ReO$_6$ octahedra got slightly distorted and they attributed this distortion to an orbital ordering of Re $t_{2g}$ electrons. Also by NPD measurement, Granado {\it et al.} have concluded that the Re 5$d$ electrons in Ca$_2$FeReO$_6$ are strongly correlated.\cite{Granado}

	The ferrimagnetism accompanied by metallic conductivity and the half-metallic DOS naturally reminds us of the colossal magnetoresistive (CMR) manganese oxides and the double exchange (DE) mechanism.\cite{CMR} Indeed, a recent theoretical study using dynamical mean field theory has shown that the DE model can explain the trend of $T_{\rm c}$ on changing Mo for Re in Sr and Ba-based double perovskites although only the Ca$_2$FeReO$_6$ case could not be explained.\cite{Chat}
	On the other hand, a different kind of models for Sr$_2$FeMoO$_6$ has been proposed by Sarma {\it et al.}\cite{Sarma,SarmaRev} and by Kanamori and Terakura.\cite{Fang,Kanamori} In those models, the hybridization between Fe 3$d$ and Mo 4$d$ states plays an important role to stabilize the FM coupling of Fe 3$d$. A strong hybridization of Fe 3$d$ and Mo 4$d$ is indeed consistent with the valence-fluctuation state of Fe$^{2.5+}$ shown by M\"ossbauer measurements on $A_2$FeMoO$_6$ (A=Ca, Sr, Ba).\cite{Linden,Nakamura} However, the model would not be able to explain the high $T_{\rm c}$ of Ca$_2$FeReO$_6$ because the reduction of hybridization strength of Ca$_2$FeReO$_6$ due to bond-angle distortion expects a lower FM coupling than Sr$_2$FeReO$_6$ or Ba$_2$FeReO$_6$ in this mechanism.\cite{Kanamori} To the best of our knowledge, no model can explain the MIT and the high ferrimagnetic $T_{\rm c}$ of Ca$_2$FeReO$_6$ at the same time.
 
	There have been several photoemission studies on the electronic structure of $A_2$FeMoO$_6$ ($A$=Sr, Ba), which have revealed many characteristic aspects of the electronic structure of the typical metallic double perovskites.\cite{Sarma,Kang,Tom,Moreo,Ray} However, Ca$_2$FeReO$_6$ has not been investigated and hence the origin of the ferrimagnetism with high $T_{\rm c}$ and the MIT of (Sr$_{\rm 1-y}$Ca$_{\rm y}$)$_2$FeReO$_6$ are not yet clarified in terms of the electronic structure. To give insight into the above problems, we have studied the electronic structure of Ca$_2$FeReO$_6$ by photoemission spectroscopy as well as LDA+$U$ band-structure calculations. Combining the electronic aspects and the ionic radius of the Re ion, we propose that the Re 5$d$ electrons are strongly correlated and therefore the number of Re 5$d$ electron has an essential importance in the MIT of this compound.

\section{Experiment and Calculation}

	Polycrystalline samples of Ca$_2$FeReO$_6$ were prepared by solid-state reaction.\cite{Kato} The site disorder was less than 5\% which would not seriously affect the microscopic electronic structure.\cite{note1} The experiments have been performed at the beamline BL-11D of the Photon Factory using a Scienta SES-200 electron analyzer.
	Surface preparation is important in photoemission measurements.\cite{McIlroy,Aiura}
	Because the samples are polycrystals, fracturing the sample in ultra-high vacuum (UHV) to expose a new surface {\it in situ} may not be always the best way to ensure a clean and representative surface, particularly at high temperatures. Hence we have done the measurements using both surface treatments.
	For low-temperature measurements at 20 K, the spectra from a fractured surface have shown better quality with respect to the intensity around -10 eV than those from a scraped surface, as is usually expected (shown in Fig.~\ref{figVBLT}); we use those spectra to compare experiment and band theory.
	For temperature-dependent measurements, on the other hand, a comparison between the two surfaces at 200 K (shown in Panel (a) of Fig.~\ref{figTdep}) confirms that the spectra from a scraped surface are generally more reliable than that from a fractured surface; we will discuss the temperature-dependence of the spectra mainly using the scraped ones. Nevertheless, we present both spectra in order to extract essential physics irrelevant to the surface treatments, particularly for the near-Fermi level ($E_{\rm F}$) region since there have been a lot of arguments on the near-$E_{\rm F}$ spectral intensity.\cite{Aiura}
	The vacuum of the measurement chamber was about 1$-$2$\times$10$^{-10}$ Torr depending upon temperatures. The surface treatments in the low-temperature measurements have been done at 20 K. In the temperature-dependent measurements, we have scraped or fractured the samples at 200 K in UHV of about 2$\times$10$^{-10}$ Torr and then run the measurements to lower temperatures in order to avoid rapid surface degradation due to outgas at heating. All the spectra have been recorded within 5$-$6 hours after each surface treatment.
	The total energy resolution was about 50$-$80 meV FWHM using 60$-$150 eV photon energies.
	We have also checked the reproducibility of the spectra. The spectral intensity was normalized by the total area of the full valence-band spectra and the near-$E_{\rm F}$ spectra were scaled according to this normalization. The $E_{\rm F}$ position has been calibrated with Au.

	Band-structure calculations have been performed with the full-potential linearized augmented plane-wave (FLAPW) method\cite{Andersen} within the local-density approximation (LDA)+$U$ scheme.\cite{Hohenberg,Anisimov} The experimental lattice parameters were those for low-temperature insulating phase, taken from Ref.~\onlinecite{Oikawa}. The plane-wave cutoff energies were 12 Ry for the wave function, and 48 Ry for the charge density and the potentials. We took 56 $k$ points in the irreducible Brillouin zone for the face centered cubic lattice.
	For effective Coulomb repulsions $U_{\rm eff}$=$U$-$J$, large values (4.0 eV for both Fe 3$d$ and Re 5$d$) have been adopted.

\section{Results}

\subsection{Band-structure calculations}

	We show the result of our LDA+$U$ band-structure calculations in Fig.~\ref{figLDAU} first. $E_{\rm F}$ was set to the top of the down-spin band for simplicity. A finite energy gap can be observed, but it is only about 30 meV.\cite{note3} It should be noted that although we used the experimental lattice parameters of the low-temperature insulating phase,\cite{Oikawa} we failed to reproduce a finite energy gap using no (namely LDA) or small $U_{\rm eff}$'s like the Sr$_2$FeMoO$_6$ case (2.0 eV and 1.0 eV for Fe and Mo, respectively).\cite{Tom} Only when we adopt a large value of 4.0 eV for $both$ Fe and Re, a finite energy gap appears. Owing to the large $U_{\rm eff}$, the up-spin band is pulled down away from $E_{\rm F}$, and the top of the up-spin band is located at -1.65 eV, which is quite deep compared with the Sr$_2$FeMoO$_6$ case, -0.8 eV.\cite{Tom} Besides, a large amount of the Fe 3$d$ up-spin states are transferred from the near-$E_{\rm F}$ region to the bottom of the valence band, -7$\sim$-8.5 eV.

	The down-spin band just below $E_{\rm F}$ is dominated by the Re 5$d$ $t_{2g\downarrow}$ and the O 2$p$ states. Because of the large $U_{\rm eff}$, the Fe 3$d$ $t_{2g\downarrow}$ states have a much smaller contribution than in the case of Sr$_2$FeMoO$_6$. 
	On the other hand, the first up-spin band below $E_{\rm F}$ is mostly due to the Fe 3$d$ $e_{g\uparrow}$ and the O 2$p$ states without any appreciable Re 5$d$ contribution. The next up-spin band from -2.5 eV to -4.5 eV has the Fe 3$d$ $t_{2g\uparrow}$ character. However, because the O 2$p$ states are the most dominant between -2.5 eV and -7 eV, the Fe $t_{2g\uparrow}$ intensity would not be obvious in photoemission spectra. 

	The eV-scale features in our calculation are essentially consistent with the recent LSDA+$U$ calculation with $U$=4.5 eV (Fe) and 1 eV (Re) by Wu.\cite{Wu} A striking difference is, however, that he has obtained a metallic DOS. Since the crystal parameters are very similar to each other,\cite{Gopala,Oikawa} the difference should come from our large $U$ for Re.
	This is supported by another theoretical calculation by Szotek {\it et al.}\cite{Szotek}  They have adopted the self-interaction corrected local spin density (SIC-LSD) scheme for the band-structure and total-energy calculations.
	The calculated DOS gives a very sharp peak of Fe 3$d$ up-spin states at about -13 eV, which is not in good agreement with ours nor Wu's. They attributed the poor agreement with Wu's calculations to a finite $U$ for Re.\cite{Szotek} Consequently, our large $U$ for Re is a key to understand the insulating DOS. We will discuss in Sec.~\ref{secDiscuss} that the large $U$ for Re is not indeed unrealistic in terms of the Re 5$d$ electron number, the ionic radius of Re ion, and the (double) perovskite-type crystal structure.

\subsection{Low-temperature spectra}

	Valence-band spectra of Ca$_2$FeReO$_6$ at 20 K taken with several photon energies are shown in Fig.~\ref{figVBLT}. Corresponding spectra from Sr$_2$FeMoO$_6$ taken with 100 eV are also shown. One can observe seven structures denoted as A to G which are essentially corresponding to the features at similar locations in the Sr$_2$FeMoO$_6$ spectra except for A at -0.5 eV and G at -10.3 eV. The intensity of G is larger in the scraped ones than in the fractured ones. This difference becomes smaller in higher photon energies and turns to be negligible at 150 eV. Such photon-energy dependence as well as its location indicates that G is mainly due to surface contaminations and/or surface aging effects. This is also in good agreement that the valence-band spectra of single crystals of a similar compound Sr$_2$FeMoO$_6$ have no intensity around -10 eV. 

	The two shoulders C (-3.8 eV) and D (-4.8 eV) are not very clear but become obvious enough in higher energies. The intense peak E (-5.8 eV) is followed by two structures F (-7.8 eV) and G (-10.3 eV). Comparing the spectra with the band-structure calculation in Fig.~\ref{figLDAU}, we can assign C and D to the Fe $t_{\rm 2g}$ up-spin states, E and F mainly to the O $2p$ states, respectively.
	The two low-energy structures A (-0.5 eV) and B (-1.5 eV) can be interpreted as the double-peak structure characteristic of the iron-based double perovskites,\cite{Kang,Tom} although they are less obvious than in Sr$_2$FeMoO$_6$. In analogy to the case of Sr$_2$FeMoO$_6$ as well as with the help of the band-structure results, these two features should be assigned to the Re 5$d$ (+Fe 3$d$) $t_{2g\downarrow}$ band and the Fe 3$d$ $e_{g\uparrow}$ band, respectively. This assignment can be tested  by a photon-energy dependent study as shown in the next figure.

	Panel (a) of Fig.~\ref{figCooper} shows near-$E_{\rm F}$ spectra of Ca$_2$FeReO$_6$ in comparison with a Sr$_2$FeMoO$_6$ spectrum taken with 100 eV. In this panel, one can see that the feature A actually consists of two fine structures A$_1$ (-0.33 eV) and A$_2$ (-0.66 eV). The location of B in Ca$_2$FeReO$_6$ is shifted away from $E_{\rm F}$ by about 0.3 eV compared with that in Sr$_2$FeMoO$_6$.
	The near-$E_{\rm F}$ spectral weight of the two features (A=A$_1$+A$_2$ and B) estimated from these spectra is shown in Panel (c) together with the theoretical spectral weight deduced from Fig.~\ref{figLDAU} and Panel (b). Here, Panel (b) shows the theoretical photoionization cross sections of the Fe 3$d$ (solid line) and Re 5$d$ (broken line) atomic states relative to the O 2$p$ atomic state (per one electron of the each states).\cite{Yeh} A$_{\rm T}$ and B$_{\rm T}$ denote the theoretical spectral weights of the feature A (from $E_{\rm F}$ to -1.5 eV) and the feature B (from -1.5 eV to -2.2 eV), respectively.\cite{note4} The integration windows for the experimental curves are -0.3$\sim$-0.8 eV (A),  -1.5$\sim$-1.9 eV (B$_1$), and -1.1$\sim$-1.9 eV (B$_2$).  For simplicity, the weight at 60 eV is set to unity for all the curves and the curves for B are shifted by 0.5.
	Panel (c) demonstrates that  A$_{\rm T}$ essentially reproduces the behavior of A, which supports our interpretation. On the other hand, the agreement between B$_{\rm T}$ and B$_1$ is not satisfactory although the high-energy end is well reproduced; the B$_{\rm T}$ monotonically increases with photon energy, mostly reflecting the increase of Fe 3$d$ cross section (see Panel (b)) while B$_1$ shows a dip behavior. This is possibly due to a remnant of the Fe 3$p$-3$d$ on-resonance enhancement at $\sim$58 eV,\cite{Kang,Ray} and if one see the region above 80 eV, B$_1$ almost monotonically increases.
	In this high photon-energy region, the discrepancy between theory and experiment becomes larger when B$_2$ is compared. Because B$_2$ includes a lower binding-energy region, this is probably a consequence that B$_2$ has the Re 5$d$ weight to some extent. In other words, the Re 5$d$ (+Fe 3$d$) $t_{2g\downarrow}$ band and the Fe 3$d$ $e_{g\uparrow}$ band are overlapping around -1 eV. This is simply because the edge of the band which accommodates two Re 5$d$ electrons should be located deeper than the edge of the band having only one Mo 4$d$ electron. Such an overlap partly explains why the double-peak structure is unclear in this compound compared with Sr$_2$FeMoO$_6$.

\subsection{Comparison between experiment and band-structure calculation}

	Figure~\ref{figVBwithLDAU} shows a comparison of the valence-band spectra at 20K with the LDA+$U$ band theory in Fig.~\ref{figLDAU}. To minimize the surface-aging effects as well as the signal from surface states, we use the 150 eV spectra to compare although the signal from the Re 5$d$ states is minimum around this energy. For a detailed comparison in the near-$E_{\rm F}$ region, we will make use of other photon energies, too. In the figure, the total DOS (gray area) includes only Fe $d$ (white area), Re $d$ (thick gray area), and O $p$ (the other area) states because the other states have negligibly small contributions. The relative intensity of the three partial DOS was fixed to that of the calculated photoionization cross-sections of each orbital.\cite{Tom,Yeh} The theoretical curve (solid curve) was obtained by broadening the cross-section-corrected total DOS with a Gaussian due to the experimental resolution as well as with an energy-dependent Lorentzian due to the lifetime effect.\cite{Tom2} The background of the experimental spectra was subtracted.

	In panel (a), one can assign the structures $\alpha$$-$$\eta$ in the theoretical curve to A$-$G in experiment. The characteristic double-peak structure (A (actually A$_1$+A$_2$) and B) is essentially reproduced in the theoretical curve as $\alpha$ (Re$-$Fe $t_{2g\downarrow}$) and $\beta$ (Fe $e_{g\uparrow}$). It is noted that $\alpha$ has also two components $\alpha_1$ and $\alpha_2$. The two structures $\gamma$ and $\delta$ are due to the Fe $t_{2g\uparrow}$ and the O 2$p$ states. The O 2$p$ contribution to $\gamma$ and $\delta$ is rather large because a large part of the Fe $t_{2g\uparrow}$ weight is redistributed to the higher binding energy region. $\varepsilon$ consists mainly of the O 2$p$ nonbonding states. Fe $t_{2g\uparrow}$ and Fe $e_{g\uparrow}$ bonding states contribute to $\zeta$ and $\eta$ to some extent, respectively.
	Although we can assign the theoretical $\alpha$$-$$\eta$ to the experimental A$-$G in this way, the agreement between theory and experiment is not satisfactory. The main part of the valence band seems to be shifted towards $E_{\rm F}$ by about 1 eV like the Sr$_2$FeMoO$_6$ case.\cite{Tom} Nevertheless, the location of $\beta$ is so deep that it is about to merge into $\gamma$. Consequently, the theoretical curve has a large intensity at $\beta$, a large portion of which is actually due to the tail of $\gamma$. The high binding energy of $\beta$ is most likely caused by the large $U_{\rm eff}$ for Fe.

	Panel (b) and (c) show detailed comparisons in the near-$E_{\rm F}$ region at 150 and 100 eV, respectively. The two fine structures A$_1$ and A$_2$ can be seen at -0.33 and -0.66 eV. The distance between them is about 0.3 eV, which is close to that between $\alpha_1$ (-0.26 eV) and $\alpha_2$ (-0.54 eV). In addition, the intensity of $\alpha_1$ relative to that of $\alpha_2$ decreases with photon energy because $\alpha_1$ has more Re 5$d$ weight than $\alpha_2$. This behavior is accordant with our observation that A$_1$ becomes less clear with increasing photon energy (see also Fig.~\ref{figCooper}). Therefore, A$_1$ and A$_2$ can be assigned to $\alpha_1$ and $\alpha_2$, respectively.
	It is obvious that even in Panel (c) the intensity of $\alpha$ is much larger than that of A in spite that its location is in good agreement with A. This sharply contrasts with the metallic Sr$_2$FeMoO$_6$ case, in which the experimental intensity of A is comparable to the theoretical one.\cite{Tom} Consequently, such a difference should be related to a strong suppression of the near-$E_{\rm F}$ spectral weight (particularly A$_1$) due to its insulating ground state although it would be partly because the present samples are polycrystals. 
	In fact, the small near-$E_{\rm F}$ intensity even above $T_{\rm MI}$ is consistent with what would be expected from the optical conductivity data which exhibited a very small Drude weight above $T_{\rm MI}$.\cite{Kato}

\subsection{Observation of MIT in near-$E_{\rm F}$ spectra}

	Temperature-dependent valence-band spectra of Ca$_2$FeReO$_6$ taken with 100 eV are shown in Panel (a) of Fig.~\ref{figTdep}. The seven structures A$-$G identified in Fig.~\ref{figVBLT} can be observed.
	Unlike the low-temperature spectra in Fig.~\ref{figVBLT}, however, the comparison between the scraped and the fractured surfaces at 200 K shows that the intensity of the feature G is considerably higher in the fractured one, indicating that a scraped surface is basically more reliable than a fractured one for the temperature-dependent measurement of this compound.

	Panel (a) shows that no spectral-weight redistribution is observed in a large ($\sim$eV) energy scale. Hence the effects of the MIT on the spectra would be observed only in the vicinity of $E_{\rm F}$. 
	Near-$E_{\rm F}$ spectra taken with 100 eV are shown in Panels (b) (scraped) and (c) (fractured). In both panels, a small but finite Fermi cut-off can be observed at 200 K. The Fermi cut-off gradually fades away with decrease in temperature and completely disappears below 100 K. This temperature is somewhat lower than the reported $T_{\rm MI}$ of 140$\sim$150 K.\cite{Kato} The energy gap in the low-temperature insulating phase is very small even at the lowest temperature. Taking into account the broadening due to the Fermi-Dirac distribution at 20 K as well as the experimental resolution, we estimate the energy gap (below $E_{\rm F}$) is about or less than 50 meV. This is comparable to the theoretical value, but the theory still gives a smaller number in spite of the large $U$ for Re 5$d$.

	In Panels (d) and (e), the near-$E_{\rm F}$ spectral weight is plotted as a function of temperature. The plots show the same behavior in spite that the typical spectral lineshapes in (a) and (b) are a little different due to the different surface treatments;
	the near-$E_{\rm F}$ weight starts decreasing from ${\sim}$175 K down to 100${\sim}$125 K, and remains nearly constant below this temperature. Note that finite weight at low temperatures is due to the finite integration window. The rather gradual decrease of the near-$E_{\rm F}$ spectral weight may be an indication of the two-phase coexistence.\cite{Granado} However, because of the very small depletion of the $E_{\rm F}$ weight across the MIT as well as the limited energy resolution, the precise determination of $T_{\rm MI}$ or an observation of the possible two-phase coexistence was beyond our experimental accuracy.

\section{Discussion}\label{secDiscuss}

	To elucidate the origin of the temperature-induced MIT in the Ca$_2$FeReO$_6$, we compare Sr$_2$FeReO$_6$ and Ca$_2$FeReO$_6$ first. As we have briefly reviewed in Sec.~\ref{secIntr}, the only difference between the two compounds is the bond angle of Fe-O-Re. Hence the most naive scenario will be that the bond-angle distortion due to Ca substitution reduces the one-electron bandwidth and leads the electron localization. 
	In reality, however, the origin of the MIT would not be only the bond-angle distortion, because Ca$_2$FeMoO$_6$, having almost the same bond-angle distortion and the crystal structure as Ca$_2$FeReO$_6$, does not show MIT.\cite{Alonso} 
	The two important differences between Ca$_2$FeReO$_6$ and Ca$_2$FeMoO$_6$ are (1) MIT occurs in Ca$_2$FeReO$_6$, while it does not in Ca$_2$FeMoO$_6$ and (2) $T_{\rm c}$ is far different to each other (Ca$_2$FeReO$_6$: 540 K, Ca$_2$FeMoO$_6$: 380 K (Ref.~\onlinecite{Alonso})). Because the crystal structure and the octahedral bond-angle distortion are almost the same, the origin of those differences cannot be attributed to crystallographic reasons. 

	Here, we focus on the ionic radius of the Re and Mo ions in stead of $A$ site cation size, which has been investigated more.\cite{Teresa} Since the valence of Fe is known as about 2.5+ from the M\"ossbauer measurements,\cite{Linden,Nakamura} we use the average ionic radius of Re$^{5.5+}$ and Mo$^{5.5+}$ to estimate the Re and Mo ionic radius, which are calculated to be 0.565 ${\rm {\AA}}$ and 0.600 ${\rm {\AA}}$, respectively.\cite{Shannon} Thus, the average ionic radius of the Re ion is actually even smaller than that of the Mo ion in spite of the difference of the 5$d$ and 4$d$ orbitals. We have also calculated the one-electron bandwidth of the hybridized Re-O or Mo-O states using the Harrison's formula.\cite{Harrison} The result is approximately the same in both systems as expected from the above consideration. Therefore, the Coulomb repulsion at the Re sites can be substantially large in Ca$_2$FeReO$_6$. 

	The key to understand this surprising consequence is the mismatch of the ionic radius of the Re ion and the (double) perovskite-type crystal structure; 
	roughly speaking, the crystal structure of $A_2BB'$O$_6$ is like a rather firm ``cage" constructed by $A$ (Ca in this case) and O in which $B$ and $B'$ (Fe and Re in this case) is located. Hence the Re ion can be smaller than the ``appropriate" ionic radius which is determined by the ionic radius ratio of Ca and O, and the Re$-$O distance is substantially larger than what is expected from a binary compound ReO$_3$. In ReO$_3$, the Re$-$O distance can be the ``appropriate" one simply because there is no other element. Thus the Re 5$d$$-$O 2$p$ hybridized bands become wide as expected, resulting in the good metallic behavior of ReO$_3$. By contrast, the large Re$-$O distance in Ca$_2$FeReO$_6$ can cause a much smaller electric conductivity although one usually thinks that a 5$d$ band must be wide.

	In this situation, another important factor is that Ca$_2$FeReO$_6$ has $two$ Re 5$d$ electrons while Ca$_2$FeMoO$_6$ has only $one$ Mo 4$d$ electron per one Re/Mo atom. This would give rise to a considerably larger effective electron-electron interaction on Re sites in Ca$_2$FeReO$_6$ than on Mo sites in Ca$_2$FeMoO$_6$, which can explain the large $U_{\rm eff}$ for Re in our LDA+$U$ calculation. The importance of the electron correlation on Re sites in the MIT of this compound has been inferred by Granado {\it et al.} first, although they did not deal with the origin of a large $U_{\rm eff}$ on Re sites in detail.\cite{Granado} Here we have demonstrated that the strong electron correlation on Re sites is actually realized in terms of the observation of MIT coupled with the LDA+$U$ band theory with a finite energy gap.
	The actual MIT of this compound is probably driven by this strong electron correlation coupled with the Jahn-Teller distortion due to the 5$d^2$ configuration both of which strongly favor the expected orbital ordering.\cite{Oikawa} Thus the origin of the temperature-induced MIT in Ca$_2$FeReO$_6$ is not just the decreasing one-electron bandwidth, but the major factor is most likely the strong electron correlation between the Re 5$d$ $t_{\rm 2g}$ electrons, which is  caused by the mismatch of the Re ionic radius and the crystal structure plus the multiple 5$d$ electrons per one Re site.

	Finally, we briefly discuss the striking difference of $T_{\rm c}$ between Ca$_2$FeReO$_6$ and Ca$_2$FeMoO$_6$ in terms of the direct overlap of Re$-$Re or Mo$-$Mo. Since the crystal structure and the bond-angle distortion of both systems are virtually identical, we can adopt the distance between the two B$^{\prime}$ site atoms and the ionic radius as measures of the amount of the direct overlap. According to the crystal structural parameters, Re$-$Re and Mo$-$Mo distance are 5.525 ${\rm {\AA}}$ and 5.522 ${\rm {\AA}}$, respectively.\cite{Alonso,Oikawa} This difference is negligibly small and indeed smaller than that of the two average ionic radii by a factor of one tenth. Thus the average ionic radius primarily determines the amount of direct overlap, and the smaller average ionic radius of Re leads the smaller Re$-$Re direct overlap than that of Mo$-$Mo. According to Chattopadhyay and Millis, a larger direct overlap decreases $T_{\rm c}$ because the ferromagnetic coupling is enhanced under the reduced direct overlap.\cite{Chat} This is in perfect agreement with our simple arguments using the ionic radius and hence further supports our arguments on the origin of MIT of this compound.

\section{Conclusion}

	We have investigated the temperature dependent electronic structure of Ca$_2$FeReO$_6$ by photoemission spectroscopy and LDA+$U$ band-structure calculations. In the valence-band spectra, we have observed the double-peak structure which is characteristic of metallic double perovskites although it was unclear compared with the Sr$_2$FeMoO$_6$ because of the larger number of the Re 5$d$ $t_{2g\downarrow}$ electrons than that of the Mo 4$d$ $t_{2g\downarrow}$ ones.
	Above $T_{\rm MI}$, the spectral weight at $E_{\rm F}$ was found to be small as expected from the optical conductivity, and completely disappeared below 100 K, forming an energy gap. The MIT occurred between 125 K and 100 K, but a finite weight at $E_{\rm F}$ due to the temperature and instrumental broadening decreases the apparent $T_{\rm MI}$. 
	Employing large $U_{\rm eff}$'s for both Fe and Re, we have reproduced the finite energy gap in the LDA+$U$ band theory. Although the most of the experimental features were interpreted with the help of the band theory, the overall agreement between theory and the valence-band spectra was not satisfactory. 
	Based on the ionic radii and the lattice parameters, we have pointed out that the effective transfer integral between Fe and Re is actually smaller than that between Fe and Mo in Ca$_2$FeMoO$_6$. This results in a substantially large electron-electron interaction on Re site (namely the Re 5$d-$Fe 3$d$ $t_{2g\downarrow}$ band), which should be the major driving force of the MIT in this compound. Finally, the very high $T_{\rm c}$ in Ca$_2$FeReO$_6$ has also been argued in connection with the Re$-$Re direct overlap.

\acknowledgements

	The authors would like to thank T. Kikuchi for technical support in the experiments, and S. Nakamura and M. Itoh for valuable discussions. Part of this work has been done under the approvals of the Photon Factory Program Advisory Committee (Proposal Nos. 00G011, 02G015, 02G016). This work was supported by a Grant-in-Aid for Scientific Research from the Japanese Ministry of Education, Culture, Sports, Science, and Technology.



\begin{figure}
  \caption{(Color online) Total and partial density of states of insulating Ca$_2$FeReO$_6$ calculated with LDA+$U$ method. $U_{\rm eff}$ is 4.0 eV for both Fe 3$d$ and Re 5$d$.}
  \label{figLDAU}
\end{figure}

\begin{figure}
  \caption{Valence-band photoemission spectra of Ca$_2$FeReO$_6$ at 20 K taken with several photon energies. Solid (dotted) lines are from a fractured (scraped) surface. Spectra of Sr$_2$FeMoO$_6$ (20 K, 100 eV) are also shown for comparison at the bottom.}
  \label{figVBLT}
\end{figure}

\begin{figure}
  \caption{
(a) Near-$E_{\rm F}$ photoemission spectra from a fractured surface of Ca$_2$FeReO$_6$ at 20 K taken with several photon energies.
(b) Theoretical photoionization cross sections of the Fe 3$d$ (solid line) and Re 5$d$ (broken line) atomic states relative to the O 2$p$ atomic state.\protect\cite{Yeh} The Re 5$d$ shows the broad Cooper minimum around 150 eV.
(c) Theoretical near-$E_{\rm F}$ spectral weight of the feature A and B (A$_{\rm T}$ and B$_{\rm T}$) compared with the experiment (A, B$_1$, and B$_2$) as functions of photon energy. Photoionization cross section of the Fe 3$d$, Re 5$d$ and O 2$p$ states\protect\cite{Yeh} are taken into accounted in A$_{\rm T}$ and B$_{\rm T}$. The integration windows are given in the text.}
  \label{figCooper}
\end{figure}

\begin{figure}
  \caption{(a) valence-band photoemission spectra of Ca$_2$FeReO$_6$ at 20 K (circles) taken with 150 eV compared with the theoretical curve (solid curve). (b) and (c) Near-$E_{\rm F}$ spectra at 20 K taken with 150 eV (panel (b)) and 100 eV (panel (c)) compared with the band theory. All the experimental backgrounds have been subtracted.}
  \label{figVBwithLDAU}
\end{figure}

\begin{figure}
  \caption{Temperature-dependent photoemission spectra of Ca$_2$FeReO$_6$ in a very near-$E_{\rm F}$ region taken with 100 eV from (a) a scraped and (b) a fractured surface. $E_{\rm F}$ spectral weight estimated by an integrated intensity from -0.1 to 0.1 eV is plotted as a function of temperature in Panel (c) (scraped) and (d) (fractured). The broken lines are guides to the eye. Hutched area shows the temperature range in which the metallic and insulating phases may coexist.\protect\cite{Granado}}
  \label{figTdep}
\end{figure}

\end{multicols}


\begin{references}
\bibitem{note1}In our previous paper, we have mentioned that the site disorder would have only minor effects on the electronic structure of Sr$_2$FeMoO$_6$.\protect\cite{Tom} The argument should also hold in the present case better, in which the site disorder is less than half of that in Sr$_2$FeMoO$_6$.
\bibitem{note3}Hence our definition of $E_{\rm F}$ does not have any serious effects when the theory is compared with the experiment.
\bibitem{note4}We did not use the simulation curve presented in Fig.~\ref{figVBwithLDAU} for this estimation because a large part of the spectral weight at B ($\beta$) is due to the tail of the feature C ($\gamma$) as shown in Fig.~\ref{figVBwithLDAU}.
\end{references}
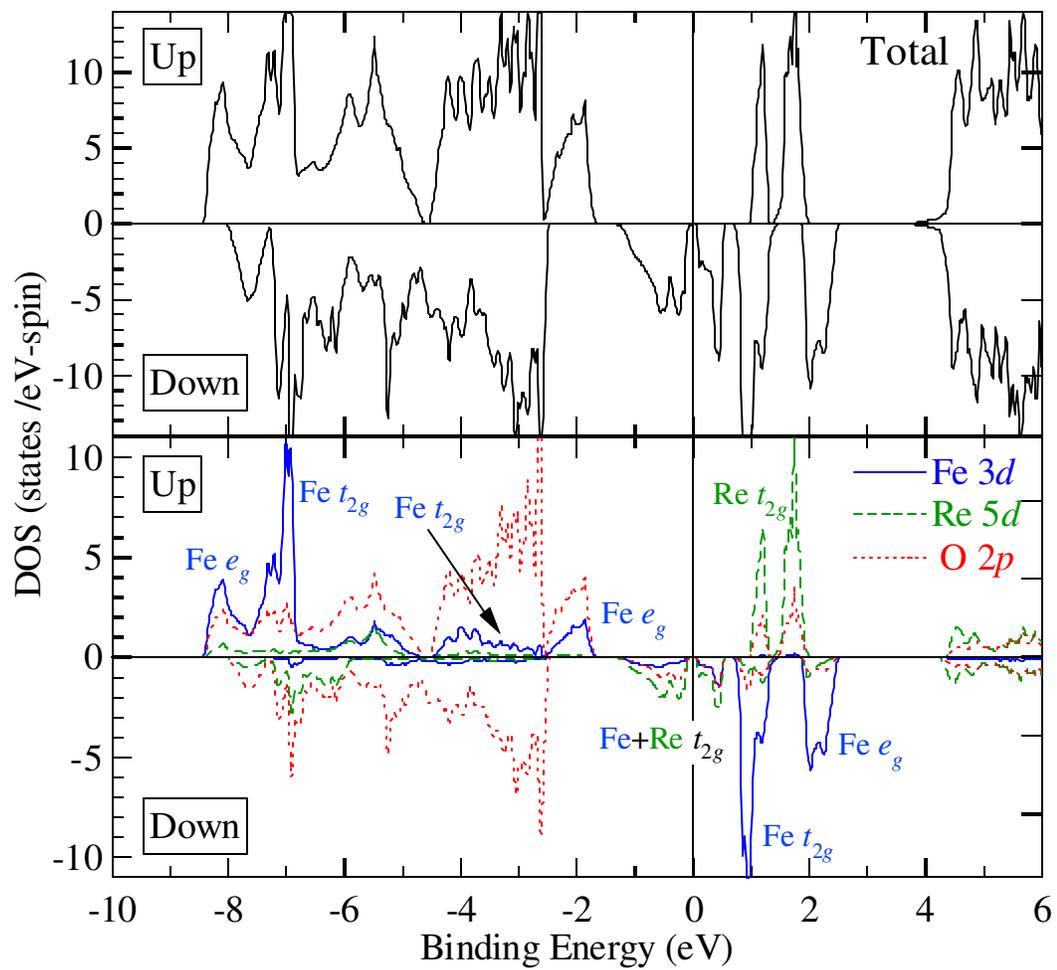
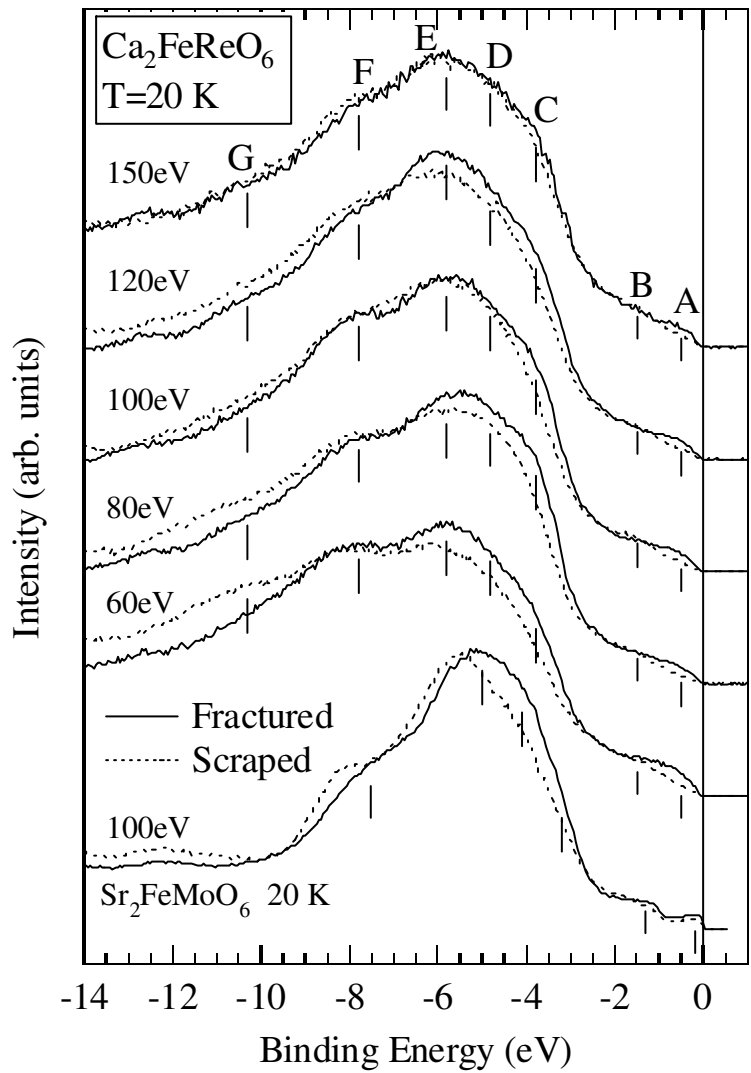
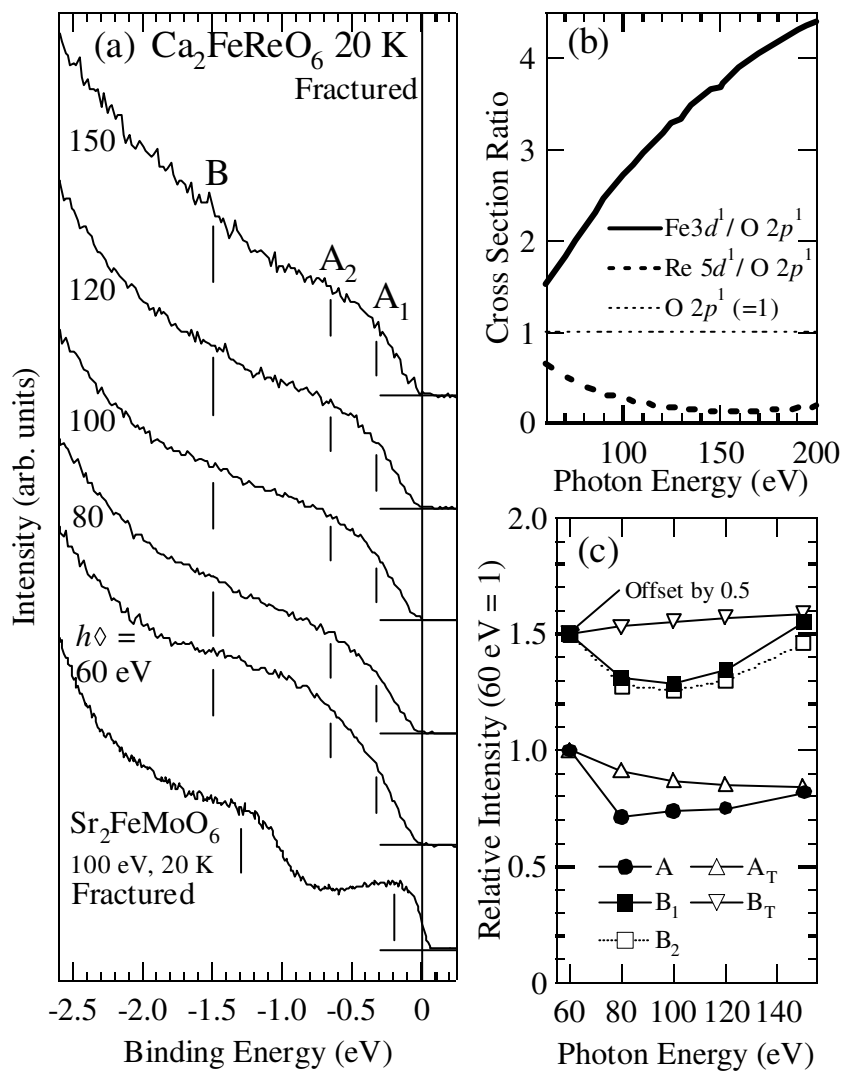
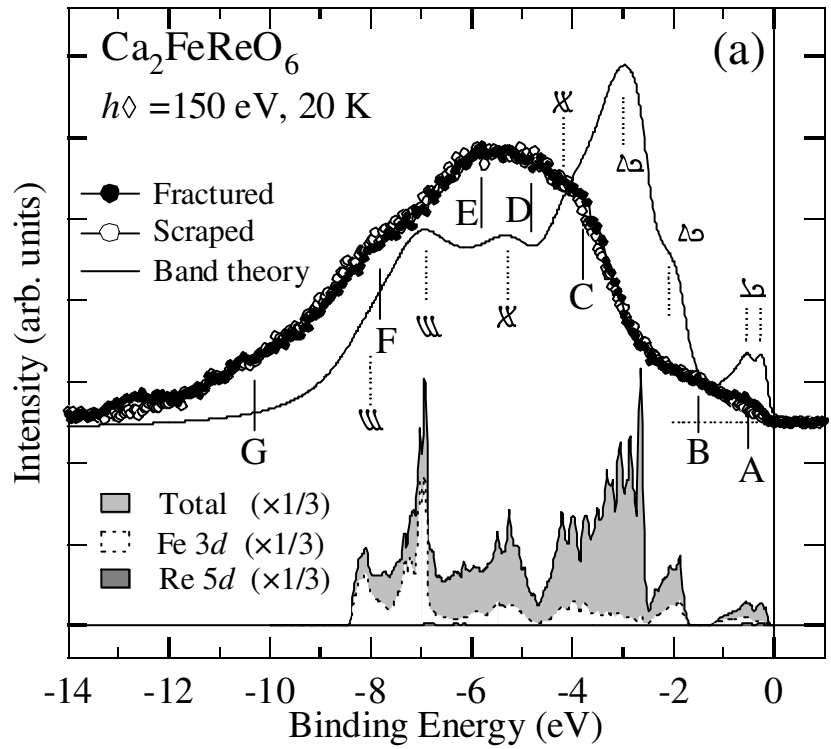
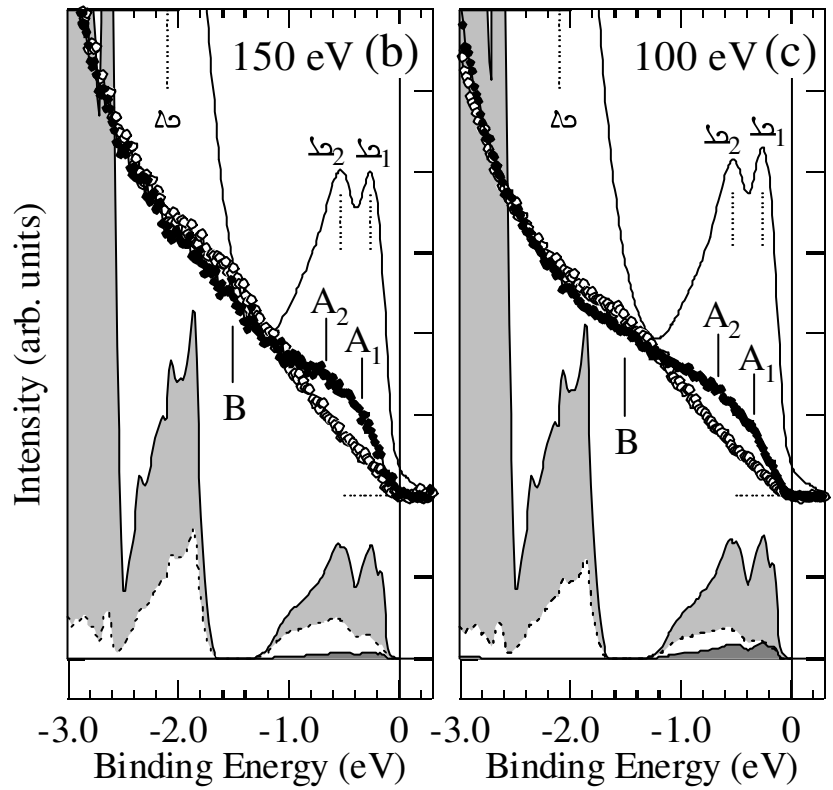
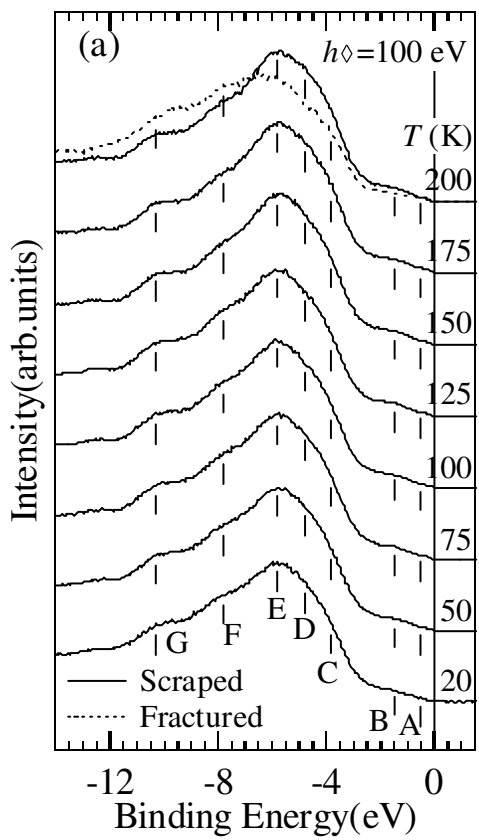
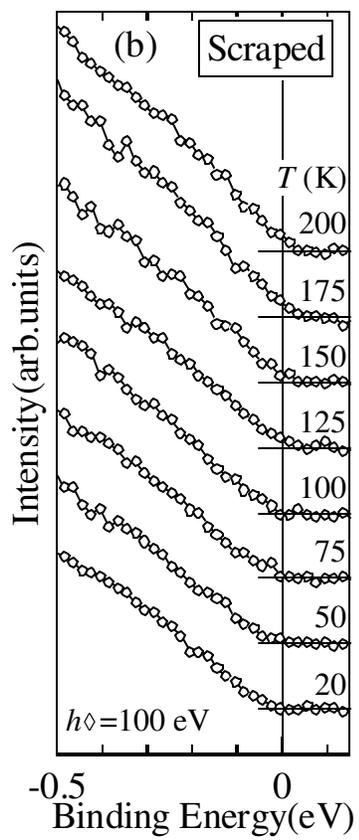
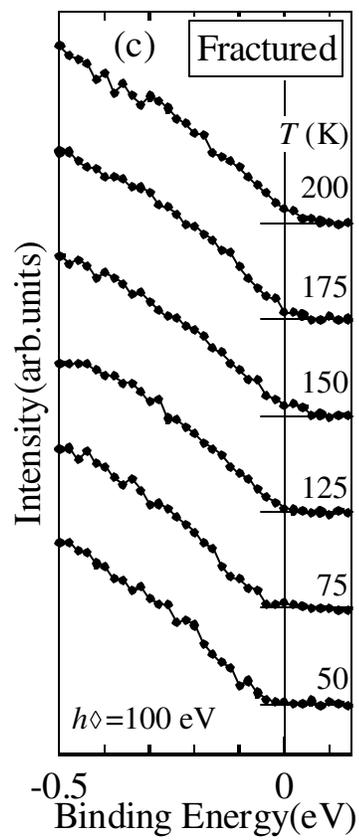
\end{document}